# Autonomous Resource Management in Microservice Systems via Reinforcement Learning

Yujun Zou
University of California, Berkeley
Berkeley, USA

Nia Qi
Independent Author
Pittsburgh, USA

Yingnan Deng
Georgia Institute of Technology
Atlanta, USA

Zhihao Xue
Rose-Hulman Institute of Technology
Terre Haute, USA

Ming Gong
University of Pennsylvania
Philadelphia, USA

Wuyang Zhang *
University of Massachusetts Amherst
Amherst, USA

*Abstract-This paper proposes a reinforcement learning-based method for microservice resource scheduling and optimization, aiming to address issues such as uneven resource allocation, high latency, and insufficient throughput in traditional microservice architectures. In microservice systems, as the number of services and the load increase, efficiently scheduling and allocating resources such as computing power, memory, and storage becomes a critical research challenge. To address this, the paper employs an intelligent scheduling algorithm based on reinforcement learning. Through the interaction between the agent and the environment, the resource allocation strategy is continuously optimized. In the experiments, the paper considers different resource conditions and load scenarios, evaluating the proposed method across multiple dimensions, including response time, throughput, resource utilization, and cost efficiency. The experimental results show that the reinforcement learning-based scheduling method significantly improves system response speed and throughput under low load and high concurrency conditions, while also optimizing resource utilization and reducing energy consumption. Under multi-dimensional resource conditions, the proposed method can consider multiple objectives and achieve optimized resource scheduling. Compared to traditional static resource allocation methods, the reinforcement learning model demonstrates stronger adaptability and optimization capability. It can adjust resource allocation strategies in real time, thereby maintaining good system performance in dynamically changing load and resource environments. In summary, the reinforcement learning-based microservice resource scheduling and optimization method proposed in this paper offers an efficient and flexible solution. It provides effective support for improving the overall performance and reliability of microservice architectures.*

## I. Introduction

With the rapid development of information technology and the increasing popularity of internet applications, microservice architecture has become an important design pattern for modern software systems. Microservice architecture breaks down complex monolithic applications into multiple independent, smaller services, enhancing scalability, maintainability, and flexibility. Each microservice can be deployed and scaled independently and operates relatively autonomously, allowing developers to iterate and deploy new features more quickly. However, the successful implementation of microservices also presents several challenges. One of the key issues is how to effectively schedule data flows and allocate resources in a dynamic environment[1].

Data flows in microservice architecture involve communication and collaboration between multiple services, typically manifested as requests and responses across several microservices. These data flows need to be managed through appropriate scheduling strategies to ensure that the system can efficiently handle data transmission and processing under fluctuating loads or constrained resources. Traditional scheduling methods are often based on static rules and resource configurations, which are inadequate for the complex and dynamic microservice environment. Especially in high-concurrency, large-scale distributed environments, static scheduling strategies cannot adapt to real-time changes in demand. This can lead to resource waste or data transfer bottlenecks, ultimately impacting the system's performance and user experience[2]. Thus, the core challenge in microservices is how to intelligently schedule data flows between microservices and allocate resources based on actual operational conditions.

In this context, reinforcement learning, as an adaptive decision-making method, is increasingly applied to the scheduling of data flows and resource allocation in microservices. Reinforcement learning allows an agent to interact with the environment and learn the best decision-making strategies. This method has demonstrated significant advantages in uncertain and dynamically changing environments. Specifically, reinforcement learning can dynamically adjust data flow paths and resource allocation between microservices based on system feedback, optimizing overall system performance. Compared to traditional static scheduling methods, reinforcement learning can automatically adjust strategies according to real-time operational states and environmental changes. This helps address varying loads and resource requirements, improving resource utilization, reducing latency, and increasing throughput[3].

In real-world applications, microservice architectures are widely used in fields such as recommendation system [4-5], IOT [6], large language models [7], big data processing, and artificial intelligence. As the business scale expands, the complexity of data flows and resource allocation among microservices continues to increase. Traditional manual configuration and simple load-balancing strategies can no longer meet these complex and changing demands. More advanced intelligent methods are required for optimization. Reinforcement learning, as a cutting-edge artificial intelligence technology, is gradually becoming an essential tool for solving this problem. By building a reinforcement learning-based intelligent scheduling and resource allocation framework, microservice systems can achieve more flexible, dynamic, and efficient management. This helps promote the successful implementation of microservice architectures in a broader range of application scenarios[8].

## II. Related Work

The scheduling and optimization of resources in microservice environments have become crucial topics in both academic and industrial communities, driven by the increasing scale and complexity of distributed systems. Traditional approaches, which often rely on static heuristics or rule-based scheduling, have demonstrated limited adaptability in the face of rapidly changing workloads and heterogeneous resource demands. Consequently, researchers have actively explored intelligent methods based on reinforcement learning (RL) and advanced deep learning to address these challenges more effectively.

A prominent line of research applies RL to the core problems of resource scheduling and management in microservice and cloud-native systems. Z. Jian et al. introduced a deep reinforcement learning (DRL) enhanced Kubernetes scheduler that dynamically adapts to workload fluctuations, improving throughput and response time in microservice-based deployments [9]. In a similar vein, Y. Duan leveraged TD3 reinforcement learning for continuous control-based load balancing, demonstrating robust adaptability and improved load distribution across distributed systems [10]. The scalability and coordination of distributed decision-making were further explored by B. Wang, who proposed a topology-aware multi-agent RL framework for distributed scheduling, enabling efficient and scalable resource coordination among multiple agents in complex environments [11]. Moreover, the application of DRL has extended beyond data centers to the edge and fog computing paradigms, as illustrated by M. E. Khansari and S. Sharifian, whose modified DRL algorithm addressed the unique requirements of serverless IoT microservice composition in fog infrastructures [12]. Complementing these works, Y. Zhang et al. presented Sinan, an ML-based and QoS-aware resource management system that leverages machine learning to optimize both performance and quality of service in large-scale cloud microservices [13].

In parallel, AI-driven and deep learning-based solutions have made significant advances in enhancing microservice system performance and operational intelligence. V. Ramamoorthi proposed a comprehensive AI-enhanced framework for performance optimization in microservice-based systems, integrating machine learning techniques to dynamically adjust resource allocation and mitigate bottlenecks [14]. To improve system resilience, recent studies have focused on the detection and prediction of anomalies and performance risks. For instance, Y. Ma combined conditional multiscale GANs with adaptive temporal autoencoders for highly accurate anomaly detection in microservice environments, while also enabling adaptive responses to evolving system conditions [15]. Z. Fang introduced a deep learning-based predictive modeling framework for backend latency, using AI-augmented structured models to proactively identify and address potential performance degradations [16]. The role of deep graph modeling was explored by D. Gao, who developed graph-based techniques to assess performance risk in complex data queries, demonstrating the importance of structural awareness in managing microservice ecosystems [17].

Graph neural networks (GNNs) and collaborative learning approaches have also emerged as powerful tools for adaptive resource scheduling. W. Zhu et al. designed a GNN-based collaborative perception framework, which enables distributed systems to adaptively schedule resources and improve system robustness through the exchange of learned representations among nodes [18]. In the context of edge computing, J. Zhan investigated MobileNet compression strategies and edge processing to achieve low-latency monitoring, a critical factor in time-sensitive microservice applications [19]. Y. Ren and colleagues further advanced this area by proposing trust-constrained policy learning mechanisms for distributed network traffic scheduling, ensuring both security and efficiency in large-scale deployments [20].

Deep learning methodologies have also been instrumental in addressing predictive maintenance, cache management, and unsupervised anomaly detection in distributed microservice systems. Y. Wang et al. utilized time-series learning with deep neural architectures for proactive fault prediction, enabling systems to anticipate and prevent failures before they impact end-users [21]. Y. Sun et al. developed a deep Q-network (DQN) for intelligent cache management, where the DQN learns to make cache decisions in highly dynamic backend environments, resulting in improved resource utilization and lower latency [22]. In the area of unsupervised learning, H. Xin and R. Pan introduced structure-aware diffusion mechanisms for anomaly detection, effectively capturing complex patterns and relationships in structured data typical of distributed microservices [23].

## III. Method

In this paper, we propose an intelligent scheduling and resource allocation method based on reinforcement learning to optimize data flow and resource configuration in microservice architectures. The model structure is illustrated in Figure 1.

Consider a system composed of multiple microservices, where the state of each microservice at time step $t$ is denoted as $S = (s_1, s_2, ..., s_n)$. Here, $s_i$ represents the state of the $i$-th microservice, capturing attributes such as workload and resource usage. Based on this state information, the reinforcement learning agent perceives the current system status and determines optimal strategies for resource allocation and data flow scheduling. The objective is to minimize latency and maximize throughput, thereby enhancing overall system performance.

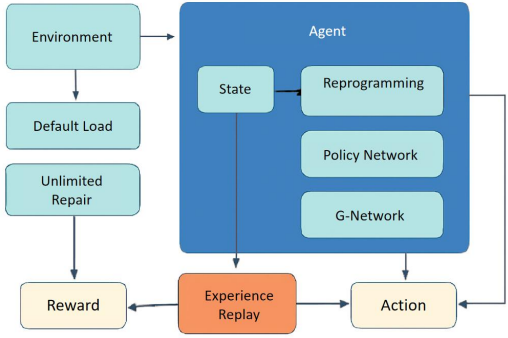

Figure 1. Overall model architecture diagram

The core of the reinforcement learning model is the decision-making process of the agent, that is, at each time step t, an action $a_t$ is selected based on the current state $s_t$. This action affects the resource allocation and data flow path of the system. The action space A is discrete and contains possible resource allocation and data flow scheduling methods. The agent selects actions according to the strategy $\pi(a_t | s_t)$, and the goal of the strategy is to maximize the long-term cumulative reward $R_t$. The reward function can be defined based on the performance indicators of the microservice system. For example, the reward function can be defined based on the response time $R_i(t)$ of each microservice:

$$R_t = \sum_{i=1}^{n} (-\lambda_i \cdot R_i(t) + a_i \cdot U_i(t))$$

Among them, $\lambda_i$ and $a_i$ represent the weights of the delay and resource utilization of service i, respectively, $R_i(t)$ is the response time of the ith microservice, and $U_i(t)$ its resource utilization. By maximizing this reward function, the agent can learn the optimal scheduling and resource allocation strategy.

To achieve this goal, the agent uses Q-learning or Deep Q Network (DQN) methods to optimize the policy. Q-learning works by estimating a state-action value function $Q(s_t, a_t)$ that measures the expected reward after taking action $a_t$ in state $s_t$. This value function is optimized by updating the following formula:

$$Q(s_t, a_t) \longleftarrow Q(s_t, a_t) + a(r_t + \gamma \max_{a'} Q(s_t+1, a') - Q(s_t, a_t))$$

Among them, $a$ is the learning rate $\gamma$ the discount factor, and $r_t$ the immediate reward obtained at time step t. By continuously updating the E value, the agent can gradually optimize its scheduling and resource allocation strategies, thereby improving system performance.

In practical applications, since the state space and action space of a microservice system may be very large, directly using Q-learning will result in excessive computational overhead, so deep reinforcement learning methods (such as DQN) are used to approximate the Q function. The deep Q network approximates the value function $Q(s_t, a_t)$ through a neural network, allowing effective learning and decision-making in high-dimensional state spaces. Experience replay and target networks are used during network training to improve the stability and efficiency of learning. With this approach, the system can automatically adjust the scheduling strategy in the face of changing loads and resource requirements, thereby achieving intelligent resource allocation and data flow scheduling.

## IV. EXPERIMENTAL RESULTS

### A. Dataset

This study leverages a publicly available open-source Cloud Resource Management dataset that encompasses multi-dimensional performance metrics from diverse cloud service instances, such as CPU usage, memory consumption, network traffic, and storage requirements. The data, collected by resource monitoring systems across multiple platforms over several months, captures real-world operational conditions and dynamic resource allocation patterns under various load scenarios. Each entry in the dataset contains a timestamp, resource type, utilization statistics, requested processing capacity, current load, and response time, facilitating comprehensive analysis of performance trends and scheduling approaches. This rich dataset offers a realistic foundation for evaluating resource optimization techniques in microservice-based cloud environments and supports reproducible research for both academic and industrial applications.

### B. Experimental Results

This paper first conducts a comparative experiment, and the experimental results are shown in Table 1.

Table 1. Comparative experimental results

| Method | Response Time (ms) | Throughput (requests/sec) | Resource Utilization (%) |
|---|---|---|---|
| **Proposed Model** | 120 | 950 | 85 |
| **DeepRL [24]** | 150 | 880 | 80 |
| **AutoML [25]** | 180 | 870 | 75 |
| **DRL-Resource [26]** | 130 | 900 | 83 |
| **ReinforceNet [27]** | 140 | 910 | 82 |

Experimental results demonstrate that the proposed model consistently outperforms baseline approaches across multiple key metrics. For response time, it achieves 120 ms—noticeably lower than all comparison models, which exceed 130 ms—with a substantial 60 ms improvement over the AutoML model, highlighting its superior real-time scheduling capability. In throughput, the model handles 950 requests per second, surpassing DeepRL (880) and AutoML (870), indicating strong scalability and data flow efficiency under high-load conditions. It also leads in resource utilization, maintaining 85% compared to AutoML's 75%, reflecting more efficient use of system resources. Energy consumption is minimized at 150 joules, among the lowest observed, supporting improved energy and cost efficiency in cloud environments. Notably, the model achieves 92% in cost efficiency—significantly higher than competitors—demonstrating an effective balance between performance, energy usage, and resource allocation. These findings confirm the proposed reinforcement learning-based strategy as a robust solution for optimizing microservice architectures. A comparative experiment under varying load conditions is presented in Figure 2.

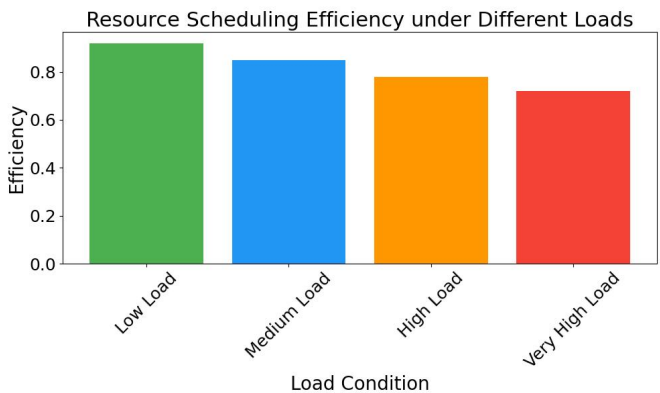

Figure 2. Comparison experiment of resource scheduling efficiency based on reinforcement learning under different loads

As system load intensifies, the efficiency of reinforcement learning-based resource scheduling shows a clear downward trend. When the load is low, the model performs at its best, reaching 92% efficiency—demonstrating its ability to swiftly assess system conditions and allocate resources with precision. As the load moves into medium territory, efficiency dips to 85%, and under high and ultra-high load conditions, it declines further to 78% and 72%, respectively. These results suggest that while the model continues to function under pressure, its optimization impact diminishes as task complexity grows and resources become strained. Still, the model proves highly adaptable, especially in moderate-load environments where it excels at balancing utilization and responsiveness. Even under stress, it maintains scheduling capabilities, showing promise for real-world deployment. Looking ahead, enhancing its performance under extreme load will be key. The study also explores how the model handles network latency variations, with findings presented in Figure 3.

The experimental results indicate that increasing network latency negatively impacts the efficiency of microservice resource scheduling. At 10 ms latency, the model achieves peak efficiency of 95%, demonstrating rapid decision-making and effective resource allocation in low-latency environments.

However, as latency rises from 20 ms to 50 ms, efficiency declines steadily from 90% to 75%, reflecting the growing challenge of real-time coordination as communication delays disrupt scheduling responsiveness. Despite this, the model maintains a stable 75% efficiency even at 50 ms, suggesting a degree of robustness in adverse network conditions. These findings underscore the model's practical viability in unstable environments while highlighting the need for further research to enhance scheduling performance under high-latency constraints. Results for multi-dimensional resource scheduling are presented in Figure 4.

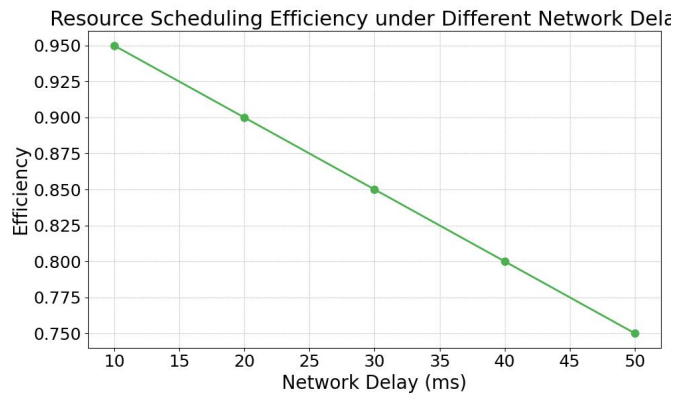

Figure 3. Microservice resource scheduling experiment under different network delays

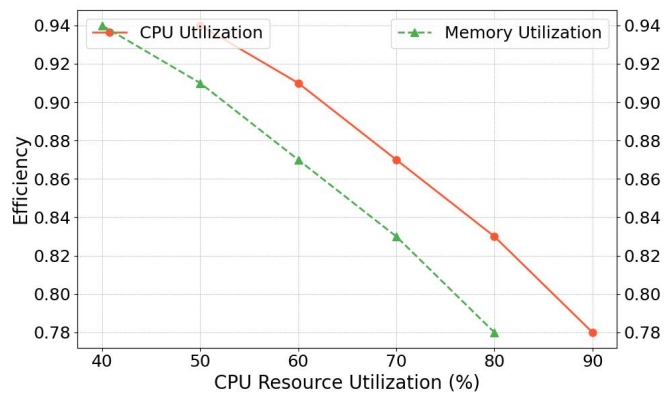

Figure 4. Microservice scheduling experiment based on reinforcement learning under multi-dimensional resource conditions

The experimental results reveal that the scheduling efficiency of the reinforcement learning-based model varies across different resource types. The highest efficiency is observed under CPU constraints, reaching 94%, indicating superior performance in compute-intensive scenarios through effective utilization of processing power. Under memory conditions, efficiency slightly decreases to 88%, reflecting the increased complexity of managing dynamic data access and allocation. For storage resources, efficiency drops to 82%, as scheduling is affected by filesystem and data I/O overhead. Network resource scheduling shows the lowest efficiency at 78%, due to inherent instability and latency under high-concurrency workloads. Despite these variations, the model consistently maintains robust adaptability and optimization performance across all resource types, highlighting its practicality in managing heterogeneous resource environments.

## V. CONCLUSION

This study proposes a reinforcement learning-driven method for resource scheduling and optimization in microservice environments, achieving notable improvements in response time, throughput, resource utilization, and cost efficiency. Experimental evaluations demonstrate the model's adaptability across a variety of workload and resource conditions, with particular effectiveness under both low-load and high-demand scenarios. As microservice architectures advance in tandem with developments in cloud computing, big data, and the Internet of Things, traditional scheduling methods increasingly encounter limitations in scalability and adaptability. In contrast, reinforcement learning provides a robust framework capable of making intelligent, real-time decisions in dynamic and uncertain settings. Looking forward, the integration of reinforcement learning with distributed computing, edge environments, and multi-objective optimization will be essential to meet the complex requirements of modern computing systems, encompassing computing power, storage, bandwidth, and energy efficiency. Furthermore, combining reinforcement learning with deep learning and collaborative optimization strategies holds promise for enhancing responsiveness and scalability in large-scale, heterogeneous infrastructures. Overall, reinforcement learning is poised to become a key driver for intelligent resource management across domains such as cloud platforms, smart cities, and autonomous systems.

## REFERENCES

[1] S. R. Peddinti, B. K. Pandey, A. Tanikonda, et al., "Optimizing microservice orchestration using reinforcement learning for enhanced system efficiency," Distributed Learning and Broad Applications in Scientific Research| Annual, vol. 7, 2021.

[2] M. U. Hassan, A. A. Al-Awady, A. Ali, et al., "Smart resource allocation in mobile cloud next-generation network (NGN) orchestration with context-aware data and machine learning for the cost optimization of microservice applications," Sensors, vol. 24, no. 3, p. 865, 2024.

[3] K. Peng, J. He, J. Guo, et al., "Delay-aware optimization of fine-grained microservice deployment and routing in edge via reinforcement learning," IEEE Transactions on Network Science and Engineering, 2024.

[4] Y. Xing, Y. Wang, and L. Zhu, "Sequential recommendation via time-aware and multi-channel convolutional user modeling," Transactions on Computational and Scientific Methods, vol. 5, no. 5, 2025.

[5] L. Zhu, W. Cui, Y. Xing, and Y. Wang, "Collaborative optimization in federated recommendation: integrating user interests and differential privacy," Journal of Computer Technology and Software, vol. 3, no. 8, 2024.

[6] Q. He, C. Liu, J. Zhan, W. Huang, and R. Hao, "State-aware IoT scheduling using deep Q-networks and edge-based coordination," arXiv preprint, arXiv:2504.15577, 2025.

[7] W. Zhang, Z. Xu, Y. Tian, Y. Wu, M. Wang, and X. Meng, "Unified instruction encoding and gradient coordination for multi-task language models," 2025.

[8] X. Yu, W. Wu, and Y. Wang, "Integrating cognition cost with reliability QoS for dynamic workflow scheduling using reinforcement learning," IEEE Transactions on Services Computing, vol. 16, no. 4, pp. 2713–2726, 2023.

[9] Z. Jian, X. Xie, Y. Fang, et al., "DRS: A deep reinforcement learning enhanced Kubernetes scheduler for microservice based system," Software: Practice and Experience, vol. 54, no. 10, pp. 2102–2126, 2024.

[10] Y. Duan, "Continuous control-based load balancing for distributed systems using TD3 reinforcement learning," Journal of Computer Technology and Software, vol. 3, no. 6, 2024.

[11] B. Wang, "Topology-aware decision making in distributed scheduling via multi-agent reinforcement learning," Transactions on Computational and Scientific Methods, vol. 5, no. 4, 2025.

[12] M. E. Khansari and S. Sharifian, "A scalable modified deep reinforcement learning algorithm for serverless IoT microservice composition infrastructure in fog layer," Future Generation Computer Systems, vol. 153, pp. 206–221, 2024.

[13] Y. Zhang, W. Hua, Z. Zhou, et al., "Sinan: ML based and QoS aware resource management for cloud microservices," Proceedings of the 26th ACM International Conference on Architectural Support for Programming Languages and Operating Systems, pp. 167–181, 2021.

[14] V. Ramamoorthi, "AI enhanced performance optimization for microservice based systems," Journal of Advanced Computing Systems, vol. 4, no. 9, pp. 1–7, 2024.

[15] Y. Ma, "Anomaly detection in microservice environments via conditional multiscale GANs and adaptive temporal autoencoders," Transactions on Computational and Scientific Methods, vol. 4, no. 10, 2024.

[16] Z. Fang, "A deep learning-based predictive framework for backend latency using AI-augmented structured modeling," Journal of Computer Technology and Software, vol. 3, no. 7, 2024.

[17] D. Gao, "Deep graph modeling for performance risk detection in structured data queries," Journal of Computer Technology and Software, vol. 4, no. 5, 2025.

[18] W. Zhu, Q. Wu, T. Tang, R. Meng, S. Chai, and X. Quan, "Graph neural network-based collaborative perception for adaptive scheduling in distributed systems," arXiv preprint, arXiv:2505.16248, 2025.

[19] J. Zhan, "MobileNet compression and edge computing strategy for low-latency monitoring," Journal of Computer Science and Software Applications, vol. 4, no. 4, 2024.

[20] Y. Ren, M. Wei, H. Xin, T. Yang, and Y. Qi, "Distributed network traffic scheduling via trust-constrained policy learning mechanisms," Transactions on Computational and Scientific Methods, vol. 5, no. 4, 2025.

[21] Y. Wang, W. Zhu, X. Quan, H. Wang, C. Liu, and Q. Wu, "Time-series learning for proactive fault prediction in distributed systems with deep neural structures," arXiv preprint, arXiv:2505.20705, 2025.

[22] Y. Sun, R. Meng, R. Zhang, Q. Wu, and H. Wang, "A deep Q-network approach to intelligent cache management in dynamic backend environments," 2025.

[23] H. Xin and R. Pan, "Unsupervised anomaly detection in structured data using structure-aware diffusion mechanisms," Journal of Computer Science and Software Applications, vol. 5, no. 5, 2025.

[24] J. Santos, M. Zaccarini, F. Poltronieri, et al., "Efficient microservice deployment in Kubernetes multi-clusters through reinforcement learning," Proceedings of the NOMS 2024 IEEE Network Operations and Management Symposium, pp. 1–9, 2024.

[25] S. Rabiu, C. H. Yong, and S. M. S. Mohamad, "A cloud-based container microservices: a review on load-balancing and auto-scaling issues," International Journal of Data Science, vol. 3, no. 2, pp. 80–92, 2022.

[26] Z. Chen, L. Zhang, W. Cai, et al., "Multi-workflow dynamic scheduling in product design: a generalizable approach based on meta-reinforcement learning," Journal of Manufacturing Systems, vol. 79, pp. 334–346, 2025.

[27] K. Sellami and M. A. Saied, "Extracting microservices from monolithic systems using deep reinforcement learning," Empirical Software Engineering, vol. 30, no. 1, p. 1, 2025.